\documentclass[prb,aps,superscriptaddress,twocolumn,showpacs,floatfix]{revtex4}
\usepackage{graphicx,amssymb, amsmath, natbib}
\usepackage{verbatim}
\usepackage{times}
\begin{document}

\newcommand{\SHOW}[1]{{\bf #1 }}
\newcommand{\CLH}[1]{{\bf #1 }}
\newcommand{\SAVE}[1] {{}}
\newcommand{\LATER}[1]{{}}
\newcommand{\prlsec}[1]{{\underline{\it #1} --- }}

\newcommand{\GG}{{\underline{G}}}  
\newcommand{\Id}{{\underline{\mathbb{I}}}}  
\newcommand{\GGO}{{\GG^0}}
\newcommand{\Sigmaa}{{\underline{\Sigma}}}  
\newcommand{\tauu}{{\underline{\tau}}}  
\newcommand{\Ree}{{\rm Re}}    
\newcommand{\Imm}{{\rm Im}}    
\newcommand{\Trk}{{{\rm Tr}_\kvec}}   
\newcommand{\coh}{{\rm coh}}
\newcommand{\calib}{{\rm cal}} 
\newcommand{\reg}{{\rm reg}} 
\newcommand{\sing}{{\rm sing}} 
\newcommand{\bos}{{\rm bos}}
\newcommand{\cosk}{{\rm cos}}
\newcommand{\BZ}{{\rm B.Z.}}
\newcommand{\saddle}{{\rm saddle}}
\newcommand{\ksaddle}{{k_s}}
\newcommand{\kvec}{{\bf{k}}}
\newcommand{\Kvec}{{\bf{K}}}
\newcommand{\qvec}{{\bf{q}}}
\newcommand{\meffcoh}{{m^*}}
\newcommand{\meffx}{{m_x}}

\newcommand{\meffy}{{m_y}}
\newcommand{\Ew}{{\omega}}   
\newcommand{\LDOS}{{DOS}}  

\title{Boson features in STM spectra of cuprate superconductors:
Weak-coupling phenomenology}

\author{Sumiran Pujari}
\affiliation{Laboratoire de Physique Th\'eorique, Universit\'e de Toulouse and CNRS, UPS (IRSAMC), F-31062 Toulouse, France;
Department of Physics,
Cornell University, Ithaca, New York 14853-2501}
\author{C. L. Henley}
\affiliation{Department of Physics,
Cornell University, Ithaca, New York 14853-2501}

\begin{abstract}
We derive the shape of the high-energy features due to a weakly coupled 
boson in cuprate superconductors, as seen experimentally in 
$Bi_2 Sr_2 Ca_1 Cu_2 O_{8+x}$ (BSCCO) by Lee \emph{et al} [Nature 442, 546 (2006)].
A simplified model is used of $d$-wave Bogoliubov quasiparticles coupled to 
Einstein oscillators with a momentum independent electron-boson coupling and
an 
analytic fitting form is derived, which allows us
(a) to extract the boson mode's frequency, and b)
to estimate the electron-boson coupling strength. 
We further calculate the maximum possible
superconducting gap due to an Einstein oscillator with the extracted electron-boson
coupling strength which is found to be less than 0.2 times of the observed
gap indicating at the observed boson's non-dominant role in the superconductivity's mechanism.
The extracted momentum-independent electron-boson coupling parameter (that we show 
\emph{a posteriori} to indeed be in the weak-coupling regime) is then to be
interpreted as an (band-structure detail dependent weighted) average over the Brillouin Zone
of the actual momentum-dependent electron-boson coupling in BSCCO. 
\end{abstract}

\pacs{74.55.+v,72.10.Fk,73.20.At,74.72.ah}


\maketitle

\SAVE{Outline of sections (1) Intro
(2) Weak-coupling model [CLH moved definitions from other sections
up into this one].  
(3) Self-energy  $\Sigmaa(\Ew)$. 
(5) Asymptotic (singular) form 
(4) \LDOS~ (boson feature). (6) Fitting (7) Discussion.
In some ways, the \LDOS~ even belongs first, but 
it depends fairly importantly on the fact that
$\Sigmaa(\Ew)$ is independent of $\kvec$, which is derived in
(3) Self energy  from the assumption that 
$g$ is independent of $\qvec$.}

\SAVE{STM spectroscopy has become a key
atomic-scale probe of correlated materials such as  high-temperature 
superconductors (HTSC).
``\LDOS'' means dI/dV,  experimentally, which is supposed
proportional to $n(\Ew)$ with an unknown coefficient.}

\section{Introduction}

\begin{figure}
\resizebox{80mm}{!}{\includegraphics{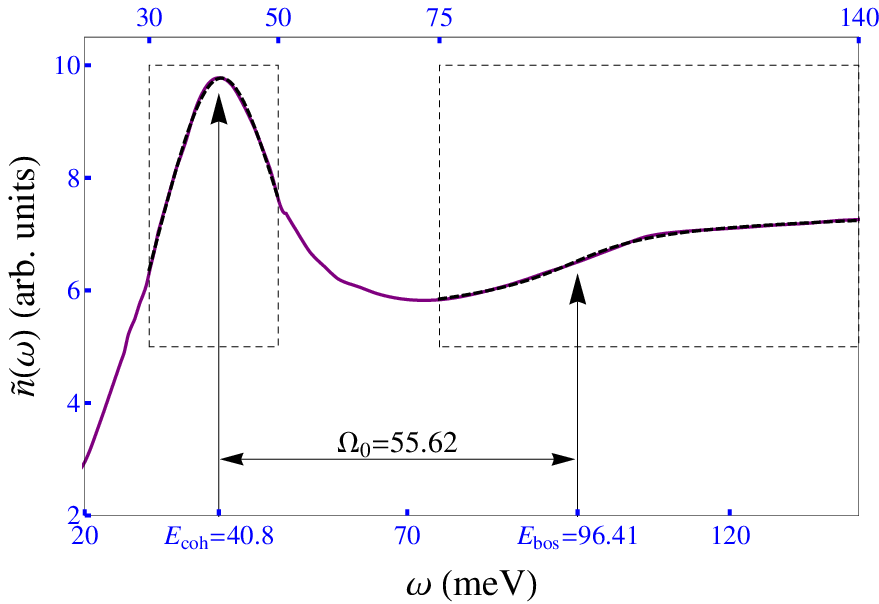}}
\caption{A typical measured STM spectrum in BSCCO (proportional to 
\LDOS~ $n(\Ew)$)
as a function of energy $E$; the data
was provided by Jacob Alldredge.
\SAVE{Numbers for this plot were obtained by 
scanning one sample data set provided by Jacob Alldredge.}
Energies of the ``coherence peak'' $E_\coh$ and boson feature $E_\bos$
are indicated.
Boxes show energy windows used to fit the analytic form (see Fig.~\ref{fig:bothfits},
below.)
}
\label{fig:Jinhoplot}
\end{figure}

Scanning tunneling microscopy (STM), applied to the superconducting
cuprate Bi$_2$Sr$_2$Ca$_1$Cu$_2$O$_{8+x}$ (BSCCO 2212) \cite{Jinho},
found a feature in the density of states (\LDOS)
at an energy well above the energy scale of the
so-called coherence peak energy (Fig.~\ref{fig:Jinhoplot}), 
and attributed it to an electron-boson coupling.
In conventional (s-wave) superconductors (e.g. Hg, Pb, Al),
such features due to electron-phonon coupling 
were known in tunneling spectra from 
superconductor-insulator-normal metal junctions~\cite{McMillan,SC-tunnel-reviews}.
\SAVE{The feature was separated from the
energy gap by the frequency of the phonons coupled
to the electrons.}
The phonon frequencies inferred from the tunneling feature
agreed with the phonon density of states inferred from
neutron scattering; furthermore, the phonon-mediated
superconducting $T_c$ and gap were correctly predicted~\cite{McMillan}
from the tunneling using the Eliashberg formalism~\cite{Eliashberg}. 
\SAVE{An excellent agreement was found of the phonon's $\alpha^2 F(\Omega)$
(product of the DOS of phonons - $F (\Omega)$ - and the (momentum-averaged) electron-phonon
interaction -$\alpha^2 (\Omega)$),
using McMillan-Rowell procedure\cite{McMillan} based on Eliashberg's
theory of strong-coupling superconductivity \cite{Eliashberg}. }
In the case of cuprates,
the mechanism for superconductivity 
is not established, and there are divergent opinions whether 
the mode observed by Lee {\it et al}
contributes to the pairing~\cite{Jinho,Carbotte2,Balatsky-localpairing,Johnston2}.
\SAVE{(Ref~\onlinecite{Jinho} hint at this but are cautious.
Zhu and Balatsky attributed the $d$-wave superconductivity to the 
bosonic mode~\cite{Balatsky-localpairing}.)}

In BSCCO, the pairing strength is highly inhomogeneous at the 
nanoscale~\cite{McElroy,inhomogeneity-refs:Pan,inhomogeneity-refs:Cren,inhomogeneity-refs:Howald,inhomogeneity-refs:Lang,inhomogeneity-refs:Fang},
as inferred from the spatial fluctuations of
the energy $E_\coh$ of the ``coherence peak''
in STM spectra (Figure~\ref{fig:Jinhoplot}).
Lee \emph{et al} discovered that the boson feature's 
energy $E_\bos$ ``floats'' with the same inhomogeneity as
$E_\coh$, namely $E_\bos= E_\coh + \hbar \Omega_0$
with a (spatially uniform) boson frequency $\Omega_0$.
To infer $E_\bos$, they identified it as the inflection point 
in DOS $n(\Ew)$ before the feature.
In this paper, we improve on this recipe by 
deriving an analytic formula for the boson feature,
starting from the simplest phenomenological model of a cuprate
and using basic RPA calculations.
Our focus here is the energy dependence rather than the spatial modulations
~\cite{Zhu1,Zhu2} of this feature.
 Prior calculations \cite{Johnston, de_Castro} 
addressed the same question of extracting $\Omega_0$ from
from the shape of the \LDOS~ of BSCCO.
Ref. \onlinecite{Johnston} uses  more
elaborate (Eliashberg) calculation, but in an entirely 
numerical framework, making the physical interpretation indirect and
the method computationally bulky to use for fitting vast number of spectra
that STM affords us with. However,
Ref. \onlinecite{Johnston} and related Ref. \onlinecite{Johnston2}
have extensively discussed the
material details about the electron-boson coupling and related form factors, that we
intentionally avoid in favour of simplicity.
\SAVE{
Did they ever actually claim in the paper that they
were devising it for real exptl fitting?}
\SAVE{(They set up an elaborate numerical Eliashberg calculation
with {\it two} boson modes. A strong mode at frequency $\approx 6 \Omega_0$
is responsible for the pairing interaction (and thus plays a role
analogous to our assumed pairing amplitude $\Delta_0$).
The second mode,  with weak coupling, is the one producing
the experimental feature.  
This work was primarily numerical and it was hard to tell 
what was physical/mathematical reason behind their conclusion.)}

We first ask just what point in the feature is to be identified as $E_\bos$:
our recipe implies a value for $\hbar\Omega_0$ in basic agreement with 
the analysis in Ref.~\onlinecite{Jinho}.
\SAVE{, and this corrected frequency is still in
disagreement with boson features seen in ARPES data.}
Secondly, we ask how can one can extract the electron-boson coupling strength; 
our results indicate it is indeed small enough that our weak-coupling
approximation is justified, and furthermore this coupling alone 
is unlikely to explain the magnitude of the observed superconducting gap.

\section{Weak-coupling Model}

We begin by setting up the simplest possible model,
taking the electron-boson coupling as a small perturbation 
to an already superconducting fermion dispersion of the standard
mean-field form (as in Ref.~\onlinecite{Zhu1, de_Castro, Eschrig-Norman, Berthod}),
and then setting up the \LDOS~
calculation within the RPA approximation.
Our analysis is agnostic as to the boson's nature,
which is sometimes argued to be magnetic~\cite{Carbotte2},
but usually considered to be an oxygen vibration,
on account of the O$^{18}$ isotope effect~\cite{Jinho}.

Our bare fermion Hamiltonian has the usual mean-field form 
\begin{equation}
{\cal H} = \sum_{\kvec, \sigma} \epsilon(\kvec) 
    c^\dagger_{\kvec, \sigma} c_{\kvec,\sigma} 
    + \Delta(\kvec) c_{\kvec,\sigma} c_{-\kvec,-\sigma} + h.c.
\label{eq:HBCS}
\end{equation}
where $\epsilon(\kvec)$ is the normal-state band dispersion, for which 
(in all numerical calculations in this paper) we adopt a six-parameter
tight-binding fit to ARPES data on BSCCO based on Ref. \onlinecite{Norman}.
The quasiparticle dispersion is then
$E(\kvec) = \sqrt{\epsilon(\kvec)^2 + \Delta(\kvec)^2}$,
where we will assume $d$-wave pairing with 
\begin{equation}
\label{eq:dwavegap}
\Delta(\kvec) \equiv \frac{\Delta_0}{2} \left[\cos(k_x)-\cos(k_y)\right].
\end{equation}
We (plausibly) approximate the bosonic mode as a dispersionless (Einstein) oscillator
at frequency  $\Omega_0$, and assume an electron-phonon coupling
    \begin{equation}
    \label{eq:el-boson}
   {\cal H}_\text{e-ph} = \frac{1}{\sqrt{N}} \sum _{\kvec,\qvec,\sigma} 
g(\qvec) c^\dagger_{\kvec+\qvec, \sigma} c_{\kvec,\sigma} 
(b_{-\qvec}+b^\dagger_\qvec)
    \end{equation}
where $b^\dagger_\qvec$ and $b_\qvec$ are the bosonic creation and
annihilation operators, and $N$ is the number of lattice sites.
For simplicity we work through the case $g(\qvec)\equiv g$;
after completing that, we will revisit the more general case
with a momentum-dependent $g(\qvec)$.

Our object, the \LDOS, is defined as the trace of the
electron term in the Green's function:
\begin{equation}
    n(\Ew) \equiv -\frac{1}{\pi} \Trk \;  \Imm \, G_{11}(\kvec,\Ew),
\label{eq:LDOS}
\end{equation}
where $\Trk\equiv a^2\int_\BZ d^2\kvec/(2\pi)^2$, and the
integral is over the Brillouin zone.
\SAVE{(CLH had wondered why only $G_{11}$ in Eq.~\eqref{eq:LDOS}
and not also $G_{22}$.
Mark Fischer explained that you get $G_{11}$ if you imagine electrons
are what tunnel, or get $G_{22}(-\Ew)$ if you imagine holes are
the particle that tunnels -- which should be physically equivalent).
In fact these are identical expressions.  So there is no
need to add both -- that would be double-counting.)}
In the $2 \times 2$ Nambu formalism, the {\it bare} Green's
function is given by
\begin{equation}
\GGO(\kvec;\Ew)^{-1} =
\left( \begin{array}{lr}
			\Ew  - \epsilon(\kvec) &  \Delta(\kvec)       \\
			\Delta(\kvec)^*    & \Ew + \epsilon(\kvec)  \\
		\end{array} \right).
\label{eq:freeSCprop}
\end{equation}
We shall henceforth use Pauli matrices $\tauu_i$,
and adopt the gauge in which $\Delta_0$ is real:
thus $\GGO= [E\Id+\epsilon(\kvec)\tauu_3+\Delta(\kvec)\tauu_1]
/[\Ew^2-E(\kvec)^2)]$.
The boson propagator has the form
\begin{equation}
D(\qvec;  \Omega ) = \frac{1}{2} \left( \frac{1}{\Omega - \Omega_0}  
- \frac{1}{\Omega + \Omega_0} \right) \equiv D(\Omega ).
\label{eq:phononprop}
\end{equation}

\section{Self-Energy and density of states due to the boson}

\SAVE{The dressed Green's function depends on the 
electron-boson coupling through the self-energy function $\Sigmaa(\Ew)$.}
The boson feature enters the \LDOS~ via the dressed Green's function, 
in the RPA approximation, 
\begin{equation}
   \GG(\kvec,\Ew)^{-1}= \GGO(\kvec,\Ew)^{-1} - \Sigmaa(\kvec,\Ew).
\label{eq:G-corr}
\end{equation}
Because $\Omega_0$ and $D( \Omega)$ were
momentum-independent, 
so is the electronic self energy,
reducing (at lowest order in $g$) to
$\Sigmaa(\kvec,\Ew)\equiv \Sigmaa(\Ew)$, where
   \begin{equation}
   \label{eq:selfenergycorr}
   \Sigmaa(\Ew)  =   g^2 \int \frac{d\Omega}{2\pi} 
      D(\Omega) \; {{\rm Tr}_\qvec } \{
\tauu_3 \; \GGO (\kvec
   -\qvec; \Ew-\Omega) \tauu_3  \}
   \end{equation}	
\SAVE{(This simplification is expedient for us as we are 
after an energy related questions.
Incidentally, so long as $g(\qvec)$ is momentum-independent and
we have an Einstein oscillator, 
\eqref{eq:selfenergycorr} 
IS the bare \LDOS $n(\Ew)$, just shifted by $\Omega_0$.)}
\SAVE{\par Eq.~\eqref{eq:selfenergycorr}
was previously been written using finite
temperature Matsubara formalism and discrete sums.
$\Sigmaa(\kvec; i \omega_n)  =  - g^2 ({T}/{N}) \sum_{\qvec, \Omega_m}$
$D(i \Omega_m) \tauu_3 \GGO (\kvec
   -\qvec; i \omega_n - i \Omega_m) \tauu_3$.
Here $\{ \omega_n \}$ and $\{ \Omega_m \}$ 
are (respectively) the fermionic and bosonic Matsubara frequencies
for temperature $T$.
That sum was then converted into  an integral for
the $T=0$ limit.
The algebra pertaining to Matubara treatment
is found in Ref.~\cite{Sumi-thesis}, Appendix C.1.)\par}
After a contour evaluation of the $\Omega$ integral Eq.~\eqref{eq:selfenergycorr} reduces to 
\SAVE{the algebra in thesis has a sign mistake in final eqn C.3}
\begin{eqnarray}
\label{eq:selfenergy}
\Sigmaa(\Ew) 
& = & \frac{g^2}{2} 
 \Trk
\Bigg\{
\frac{(\Ew+\Omega_0) \Id + \epsilon(\kvec) \tauu_3 - \Delta(\kvec) \tauu_1}{(\Ew + \Omega_0)^2 - E(\kvec)^2} 
 \nonumber  \\
& & + \frac{\Omega_0 ( \Id + \frac{\epsilon(\kvec)}{E(\kvec)} \tauu_3
- \frac{\Delta(\kvec)}{E(\kvec)} \tauu_1 )}{[\Ew - E(\kvec)]^2-\Omega_0^2} .
\Bigg\}
\end{eqnarray}
\SAVE{ 
(Note the $i$ in front is from capturing residues in Matsubara summation trick
see SP thesis, App C.1.)}
The off-diagonal ($\tauu_1$) terms in Eq.~\eqref{eq:selfenergy}
vanish, $\Sigma_{12}=\Sigma_{21}\equiv 0$, since
$\Delta_{\kvec}$ has $d$-wave symmetry (reverses sign under 
90$^\circ$ rotations).
\SAVE{(The other two terms, in $\Id$ and $\tauu_3$, 
have $s$-wave symmetry.)}

\begin{figure}
\resizebox{85mm}{!}{\includegraphics{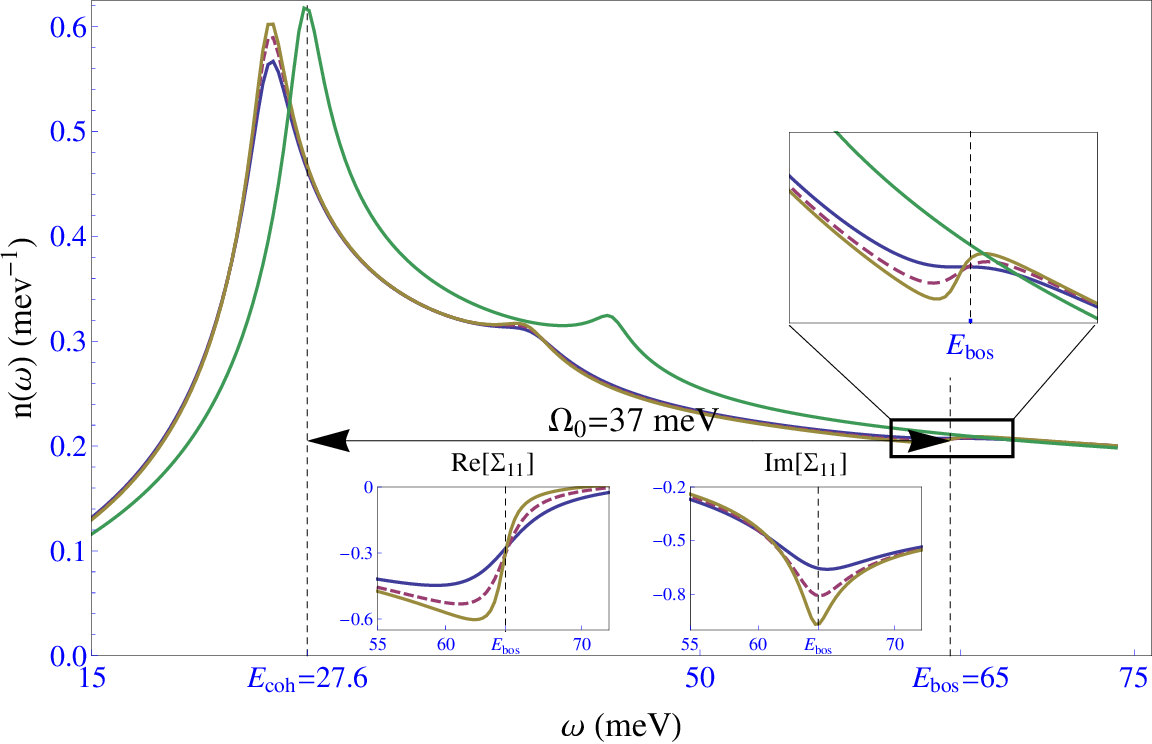}}
\caption{
[COLOR ONLINE]
Numerically computed \LDOS~ as a function of energy.
The standard six-parameter dispersion for
BSCCO was used~\cite{Norman}; we chose the pairing amplitude to be 
$\Delta_0 = 0.2|t_1| \approx 29.6$ meV,
    \SAVE{So coh. peak is at $\approx 0.185 t_1 = 27.4$ meV}
the boson energy to be $\hbar \Omega_0 = 0.25|t_1| \approx 37$ meV
and electron-boson coupling to be $g = \sqrt{0.1}|t_1| \approx 47$ meV,
where $t_1$ is the nearest-neighbor hopping.
Three different choices of damping are shown:
from sharpest to flattest, $\eta=0.005$, $0.01$ and $0.02$ 
(in units of $t_1$).
Lower inset shows real and imaginary parts of the electron self-energy
function. This has singularities at $E_\bos$ rounded by the damping.
\SAVE{The subscript is a Nambu index. The off-diagonal 
matrix elements are zero; $\Sigma_{22}(\Ew)$, although different from
$\Sigma_{11}(\Ew)$, has the same singular behavior around the feature.
$\Ree \Sigma_{11}$ shows a logarithmic singularity, as seen where
the height increases in equal steps as $\eta$ decreases by factors
of two;  $\Imm\Sigma_{11}(\Ew)$ evolves to a step in the limit $\eta\to 0$.
The boson feature energy $E_\bos$, 
offset from the ``coherence peak'' by the boson mode's frequency,
falls close to the inflection point \emph{after} the hump
(see zoom in second inset).}
}
\label{fig:BosonSelf+LDOS}
\end{figure}

\SAVE{An alternate path:
We substitute \eqref{eq:LDOS} into \eqref{eq:G-corr}, and get
$n(\Ew)  = -\frac{1}{\pi} \Imm  \Trk
\{[\Ew+\epsilon(\kvec) - \Sigma_{22}(\Ew)]/{\rm Det}(\GGO(\kvec,\Ew)) \}$
where ${\rm Det}$ means the determinant.}

We write $n(\Ew)=n_0(\Ew) + \delta n(\Ew)$, where
$n_0(\Ew)$ is tbe basic \LDOS~ in the absence of the boson coupling
[derived from \eqref{eq:freeSCprop}]
and has the well-known ``coherence peaks''
centered at energy values $\pm E_\coh$ close to $\pm \Delta_0$;
$\delta n(\Ew)$ contains contributions of order $g^2$,
in particular the boson feature.
Writing the Taylor expansion of \eqref{eq:G-corr},
\SAVE{(The intermediate equation is:
   $\delta n(\Ew)= -\frac{1}{\pi} \Imm \Trk 
    [\GG_0 \Sigmaa(\kvec,\Ew) \GG_0 ]_{11}.$ }
we extract the terms in $G_{11}$ linear in $\Sigmaa$ and thus
   \begin{eqnarray}
   \delta n(\Ew)&=& -\frac{1}{\pi} \Imm
   \Trk \left\{ 
\frac{ [\Ew+\epsilon(\kvec)]^2 \Sigma_{11}(\Ew) + |\Delta(\kvec)|^2 \Sigma_{22}(\Ew) }
  {[\Ew^2-E(\kvec)^2]^2} 
    \right\} 
     \nonumber \\
    &\equiv& \frac{1}{\pi}\Imm 
           \left\{ I_1(\Ew) \Sigma_{11}(\Ew) + I_2(\Ew) \Sigma_{22}(\Ew) \right\}.
   \label{eq:BosonLDOScorr}
   \end{eqnarray}
This is the first version of our result,
suitable for numerical fits~\cite{Jacob-new},
but requiring integrations over the zone at
each interation [for the key formulas ~\eqref{eq:selfenergy}
and ~\eqref{eq:BosonLDOScorr}.
Note that in numerical calculations,
we replace 
$\omega \to \omega + i\eta$ 
\SAVE{Sumi believes after reflection 26oct12
that the preceding is the correct recipe, and
NOT \epsilon(\kvec) \to \epsilon(\kvec) + i\eta(\kvec)$.}
in \eqref{eq:freeSCprop}, 
where $i\eta$ represents the physical quasiparticle damping
(from all sources except our boson mode), a
parameter found essential for fitting
the ``coherence peaks'' in the \LDOS~\cite{Jacob}.
\SAVE{In any case, a small imaginary part is needed to
smooth the spiky artifacts consequent on discretizing $\kvec$
but in the present context, it seems the actual dampling is larger.}
(It is easy to replace this energy-independent damping
by $\eta(\Ew)$, as used in Ref.~\cite{Jacob}.
\SAVE{The only ``numerical calculations'' we do are those
in Figure~\ref{fig:BosonSelf+LDOS}. For the values used, 
see its caption; we typically adopted $\eta$ values of order 1 meV.}
Fig.~\ref{fig:BosonSelf+LDOS} shows a 
representative numerical calculation of the self-energy function
(inset) and the resulting \LDOS.
We see a dip-hump shape, in agreement with experiment; 
$E_\bos$ falls between the dip and the hump similar to the assumption of Ref. \cite{Jinho}. 


\SAVE{There is a nice figure of $I(\Ew)$ but there is no room for it.}

\section{Asymptotic Form near $E_\bos=E_\coh + \Omega_0$}

We now extend our results to an approximate {\it analytic} formula, 
for the boson feature's shape, by treating not only the electron-boson 
coupling $g$, but also the damping $\eta$ as a small parameter:
in the limit $\eta\to 0$ the feature is a singularity 
centered at $E_\bos \equiv E_\coh + \Omega_0$.
\SAVE{
As anticipated, starting with Ref. \cite{Jinho}.
One motivation of our work is to better understand how it depends on
physical parameters, and for fast fitting.}

First recollect the origin of the familiar ``coherence peak''
in the basic \LDOS~ $n_0(\Ew)$: it is a van Hove singularity
due to the saddle points at $\kvec=(\ksaddle,\pi)$ and equivalent
momenta where the Fermi surface crosses the zone boundary.
The pertinent pole in $\GG_0$ is $\frac{1}{2}[\Ew-E(\kvec)]^{-1}\Id$;
there is no contribution from $\tauu_3$ due to the factor $\epsilon(\kvec)$
which vanishes on the Fermi surface.
It is well known that $\Trk[\Ew-E(\kvec)]$ at a saddle gives
a logarithmic singularity, so we find a singular part
\begin{equation}
\label{eq:cohsing}
n_0^\sing(\Ew) = - \frac{a^2 \meffcoh}{\pi^2} 
{\rm Re} f(\Ew - E_\coh + i \eta_\coh)
\end{equation}
with 
\begin{equation}
   f(z) \equiv \ln (\meffcoh z/4 K_x K_y).
\label{eq:f_of_z}
\end{equation}
Here $E(\kvec) \approx E_\coh + (k_x-\ksaddle)^2/2\meffx 
-(k_y-\pi/a)^2/2\meffy $ near the saddle,
$\meffcoh\equiv \sqrt{\meffx \meffy}$, 
and $K_x$, $K_y$ are cut-offs, 
representing the range of $(k_x,k_y)$ within which this expansion
is valid.
For our parameters, $1/{\meffcoh} = 94.07$meV $a^2$, 
and we take $K_x = 0.5 a^{-1}$ and $K_y = 0.06 a^{-1}$ 
for later numerical calculations.

\SAVE{
It would actually be cleaner to define CLH's function
$f(z)$ to be the divergent part of $\Trk\ [\Ew-E(\kvec)]^{-1}$.}

The self-energy $\Sigmaa(E)$ has a singularity due to the 
same saddle point, with the pole of form 
$(\frac{1}{2})^2 g^2 \Id[\Ew-E(\kvec)-\Omega_0]^{-1}$, coming 
from the second big term in \eqref{eq:selfenergy}.
Clearly, integrating over $\kvec$ gives the same logarithmic
divergence, with its argument shifted by $\Omega_0$.
Thus,
\begin{equation}
\label{eq:Sigma_sing}
\Sigmaa(\Ew+i \eta) =\text{regular terms} +
  \frac{ i g^2 a^2}{2\pi}  \meffcoh  f(\Ew - E_\bos + i \eta) \Id 
\end{equation}
with $f(z)$ from \eqref{eq:f_of_z}.
This behavior is confirmed by the inset of Fig.~\ref{fig:BosonSelf+LDOS}.

The $\Id$ dependence in \eqref{eq:Sigma_sing}
signifies that $\Sigma_{11}\approx \Sigma_{22}$ at the singularity.
Thus \eqref{eq:BosonLDOScorr} simplifies to 
\begin{equation}
\label{eq:deltaN}
   \delta n^\sing(\Ew) = \frac{1}{\pi} \Imm [ \Sigma_{11}(\Ew) I(E_\bos+i\eta)], 
\end{equation}
with $I(\Ew)\equiv [I_1(\Ew)+I_2(\Ew)]/2$ (see Eq.~\ref{eq:BosonLDOScorr}).
\SAVE{Since the self energy is independent of momentum 
[cf. Eq. \eqref{eq:energyindep}],
the self energy 
term in the numerator of momentum integral of the self energy corrected propagator  
does not get affected by the momentum integration and the integral $I(\Ew)$
provides an overall
multiplicative factor to the singularity of the self-energy 
[cf.  Eq.~\eqref{eq:BosonLDOScorr}.]}

Thus our key asymptotic result is that 
$\delta n(\Ew)$ has a \emph{logarithmic} singularity
at $E_\bos$,  rounded by the finite damping $\eta$.
The result is a linear combination of a rounded step and
a cut-off log divergence, with the exact shape
(and the location of $E_\bos$ within it) depending on 
the phase angle in $I(\Ew)\equiv|I(\Ew)|e^{i\phi_I}$,
which depends on the band structure
[cf. Eq.~\eqref{eq:deltaN}].

\SAVE{SP email 7/19/12:  $I(\Ew)$ has two roles.
(1) the signs and relative magnitudes of its real and imag. parts are
responsible for the shape of the feature (2) it is (one of) the
weighting functions that relate $g^2$ to the 
average of $|g(k)|^2$.}

For energies around the boson feature (e.g. $\Ew \approx 115$ meV), 
the rough dependence on damping is
$I(\Ew+i\eta) = 1.5 \times 10^{-5}(\eta-15)+ 0.7\times 10^{-3} i$.
Thus, the shape of $\delta n(\Ew)$ is a (comparable) combination of a
rounded {\it upwards} step from $\Ree\Sigma_{11}$ and a
rounded logarithmic hump from $\Imm \Sigma_{11}$ leading to
location of the boson mode frequency $\Ew$ before the hump
(as seen in numerics cf. Fig. \ref{fig:BosonSelf+LDOS}).

\SAVE{The logarithm function  f(z+i eta), has
a negative hump in Re f(z) and a step down in Im f(z), if
eta>0.  Feature is $\Imm[I(\Ew)* f(\Ew)]$
Then  $\Ree I(\Ew) \Imm f(\Ew)$ with 
$\Ree I<0$ should cause a step up, while
$\Imm I(\Ew) \Ree f(\Ew)$ with 
$\Imm I<0$ should cause an upwards hump.
In fact (see CLH email 10/25), an  upwards step seems
to go with physical intuition (and also, CLH thought, the
Greens function result from Johnston.)
}

\SAVE{(Note from CLH 10/25/12): Can we guess
the signs in of $I(\Ew)$ analytically, based on its double
pole singularity?}

\SAVE{It appears analytically (CLH) that $I_1(\Ew)$ and $I_2(\Ew)$ 
should both vanish as $\Delta_0^2$ as $\Delta_0 \to 0$.  
This has not been checked.}

\SAVE{The following contradicts the paragraph in thesis
footnote~\onlinecite{Sumi-thesis} has the retraction}
We can attach physical interpretations\cite{Sumi-thesis} to the real and
imaginary parts of $\Sigma_{22}(\Ew)$.  The imaginary
part represents an inelastic event in which a {\it real} 
boson excitation is created;
\SAVE{(It could equivalently be evaluated
from the matrix element for that process using Fermi's Golden rule.)}
the real part represents the quasiparticle being dressed by
{\it virtual} bosons.

Since the predicted feature includes a ``step up'',
we are in agreement with the recipe of Lee~\emph{et al} which placed
 $E_\bos$ at the inflection point
{\it before the hump} of the boson feature, motivated
by previous work on molecular vibrational features
in electron tunneling \cite{Jaklevic, Stipe}.
Refs.~\onlinecite{Johnston}, \onlinecite{de_Castro} and \onlinecite{Pasupathy}
\SAVE{(Johnston based on a numerical Eliashberg calculation)}
located $E_\bos$ even lower, at the {\it minimum of the dip} in the dip-hump feature.
As mentioned before, we also place $E_\bos$ {\it before} the hump
but more specifically in between the hump and its preceding inflection point.

We can attempt to compare our self-energy functions with those
of Ref.~\cite{Johnston} [(Figure 3(c)],  computed numerically 
from Eliashberg theory.  ${\rm Re} \Sigma_{ii}(\Ew)$ is proportional
to their ${\rm Im} Z(\omega)$ which indeed resembles a (positive) log divergence,
while ${\rm Im} \Sigma_{ii} \propto 1-Z(\omega)$ shows a rounded up step.

\SAVE{
Ref.~\onlinecite{Johnston} says (copied from our intro.)
that $E_\bos$ lies at the {\it minimum of the dip} in the dip-hump feature. 
There seems to be NO value of the phase angle $\phi_I$
for which that be the right answer? 
SP says for any value of phase angle $\phi_I$, we find $E_\bos$
lies {\it between} the two inflection points on either side of the hump.}

\section{Fitting Scheme for the Experimental Boson Feature}

In this section, we translate our asymptotic forms to a simplified fitting
scheme for our weak-coupling model and,
by applying it to the experimental spectrum in Fig.~\ref{fig:Jinhoplot},
extract the $E_\bos$ and also obtaining the electron-boson
coupling $g$ from the boson feature's amplitude \cite{E>0}.
We consider the experimental signal to be in arbitrary units
so we write it $\tilde{n}(\Ew)= \beta_\calib n(\Ew)$,
where the coefficient $\beta_\calib$ includes
unknown factors such as the STM tip set-point.  
\SAVE{Factors relating to the STM tip shape and its offset from the surface.
Specifically $\tilde{n}(\Ew)$ is the differential conductance $dI/dV$.}
As the dispersion $\epsilon(\kvec)$ is already known from ARPES~\cite{Norman},
the ``coherence peak'' is sufficiently constrained that we can calibrate 
$\beta_\calib$ from it.
We read off $E_\coh =40.8$ meV  from the peak position in Fig. \ref{fig:Jinhoplot}.
From this, using $E_\coh =E(\kvec_\saddle)$, we infer
$\Delta_0 = 44.23 \text{meV}$.
\SAVE{Of course, the maximum is not exactly $E_\coh$.
Also when we do this here, we are implicitly assuming that
the boson contribution at this energy is negligible.}

The saddle point of the quasiparticle dispersion at $\kvec_\saddle$ 
contributes a logarithmic singularity
to the \LDOS~ at the ``coherence peak'':
\begin{equation}
\label{eq:cohfit}
n_0(\Ew)  = n^\reg(\Ew) + n_0^\sing(\Ew)
\end{equation}
where $n_0^\sing(\Ew)$ is given by \eqref{eq:cohsing},
and we adopt the simplest usable form $n^\reg(\Ew)= a_\coh \Ew + b_\coh$
for the regular part, which is due mainly to $n_0(\Ew)$.
\SAVE{The regular contributions of the bosonic self energy 
around the coherence peak are also part of this background.''
There are no singular contributions to the bosonic self-energy 
near the coherence peak energy.}

\begin{table}
\begin{tabular}{|lc||lc|}
\hline
$\Delta_0$    &  44.23 meV &
        $\Omega_0$ &  56(1) meV \\
$\beta_\calib$  & 3.2(4)$\times$10$^{4}$ arb. units meV$^{-1}$&
         g & 36(16) meV \\
$\eta_\coh$ &  10.7(9) meV &
         $\eta_\bos$ &  11(2) meV \\
$a_\coh$ &  3.1(2) $\times$ 10$^{-2}$ meV$^{-2}$ &
        $a_\bos$ & 0.40(35) $\times$ 10$^{-2}$ meV$^{-2}$ \\
$b_\coh$ &  8.1(7) meV$^{-1}$ &
         $b_\bos$  & 6.9(5) meV$^{-1}$ \\
\hline
\end{tabular}
\caption{
\label{tab:bothfitresults}
Fit parameters for the ``coherence peak'' using
Eq.~\eqref{eq:cohfit} (left column) and for the boson
feature using Eq.~\eqref{eq:bosonfit} (right column).
The error-bars on the fit parameters were estimated
 by determining the parameter range where $\chi^2 \leq 2*\chi^2_{min}$,
 where $\chi^2 = \sum_i (y_i - f(x_i))^2$, $i$ is the (energy) index
 for data-points, $y_i$ and $f(x_i)$ are the experimental datum 
and the value of fitting function respectively at the $i$-th data point.
\SAVE{The arbitrary units are those of the data provided by
Alldredge et al.  It is worthwhile including this because the
error (relative to the value) is meaningful.}
}
\end{table}

\begin{figure}[h]
\begin{tabular}{cc}
a) \resizebox{38mm}{!}{\includegraphics{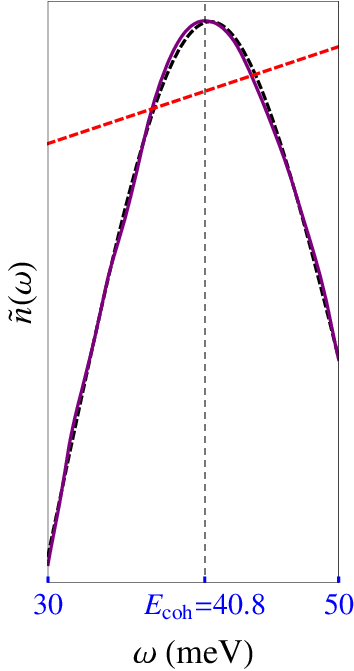}} & b) 
\resizebox{38mm}{!}{\includegraphics{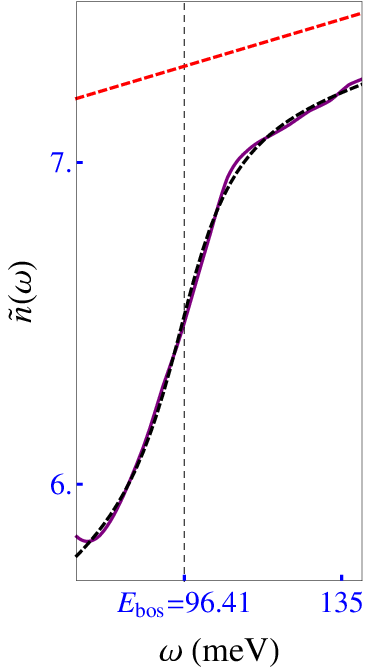}}
\end{tabular}
\caption{Fit of the experimental \LDOS~ $n(\Ew)$ to fitting forms,
using the windows of energies marked in Fig.~\ref{fig:Jinhoplot}.
(a) Fit of ``coherence peak''
to Eq.~\eqref{eq:cohfit}, using energies 30--50 meV.
(b) Fit of boson feature to Eq.~\eqref{eq:bosonfit},
using energies 80--140 meV.
}
\label{fig:bothfits}
\end{figure}

Table~\ref{tab:bothfitresults} gives
the results of the calibration fit 
to the data in Fig. \ref{fig:Jinhoplot},
using energies in (30 meV, 50 meV).
As Fig. \ref{fig:bothfits} (left panel) shows, 
the fitting is \emph{good} in this window. 
This fit gives a quasiparticle broadening
$\eta_\coh \approx 10$ meV (assumed to be constant over the 
Brillouin zone and the energy window $30-50 meV$), 
uncomfortably large in that $\eta_\coh/E_\coh \approx 1/4$.
\SAVE{Note we have used the value
of $\eta$ at the saddle point as representative of the inverse lifetime over the zone,
which might not be sensible.}
We do not know why this exceeds the result
$\eta(E_\coh \approx 40 \text{meV}) \approx 1$ meV.
fitted by Ref.~\onlinecite{Jacob} assuming a broadening
$\eta(\Ew)\propto \Ew$.
\SAVE{
The large $\eta_\coh\approx \eta _\bos$ \approx 10$. 
It is somewhat surprising that $\eta_\coh \approx \eta_\bos$,
CLH expected $\eta_\coh < \eta_\bos$.
This matters since the $\eta$ results directly affect $g$.
But remember the error bars are pretty large.}

Now we turn to the fit of the boson feature, 
using an energy window (80 meV, 140 meV) which contains the hump in 
Fig. \ref{fig:Jinhoplot},
to the fitting form implicit in Eqs.~\eqref{eq:BosonLDOScorr}
[for $I(\Ew)$], ~\eqref{eq:Sigma_sing}, and ~\eqref{eq:deltaN}:
\begin{equation}
\label{eq:bosonfit}
  n(\Ew) = n_\bos^\reg (\Ew) + \delta n^\sing(\Ew).
\end{equation}
Here we take the simplest usable form for the
regular part $n_\bos^\reg (\Ew) = a_\bos \Ew + b_\bos$,
representing $n_0(\Ew)$ 
plus regular contributions from $\Sigmaa(E)$.
Also from \eqref{eq:deltaN} we see
\begin{eqnarray}
\label{eq:deltaN-2}
\delta n^\sing(\Ew) &=&
\frac{2 i g^2 a^2 \meffcoh} {(2\pi)^4} \times \\
\times \Imm && \left[ I(\Ew + i \eta_\bos) \cdot
f(\Ew - E_\bos+ i \eta_\bos)\right] \nonumber 
\end{eqnarray}
The fitted parameters are given in Table~\ref{tab:bothfitresults}; the fit 
(Fig.~\ref{fig:bothfits}) is \emph{fairly good} in its energy window.
\SAVE{The fit procedure used the {\tt FindFit} function of the 
{\tt Mathematica} software package.}

We note that, based on the data from which Fig.~\ref{fig:Jinhoplot} is drawn,
Ref.~\cite{Jinho} identified the bosonic mode energy as $52 \pm 8$ meV,
using the inflection point before the hump, so our result
of $56 \pm 1$ meV (fitting just one \emph{typical} spectrum) is in
agreement with them.
\SAVE{(In earlier drafts, we had the wrong sign on the self-energies
which let us to think the boson frequency was after the hump, and
we thought there was a discrepancy in the frequencies.
In the corrected version, the boson frequency is indeed before the hump 
(in agreement with J. Lee et al).}
The quasiparticle damping
\SAVE{(assumed independent of energy within the fit window)}
was $\eta_\bos \approx 11 \text{meV}$.  
Thus $\eta(\kvec)/E(\kvec)\approx \eta_\bos/E_\bos \approx 0.11 \ll 1$
in the $E_\bos$ fit window,
verifying the criterion for the Bogoliubov quasiparticles to be
well-defined.

A dimensionless measure of coupling strength is the
the ratio of the logarithmic factors $f(z)$ in the
boson feature [Eq.~\eqref{eq:bosonfit}] and coherence peak 
[Eq.~\eqref{eq:cohfit}]:
\begin{equation}
\label{eq:weakcouplingratio}
\lambda_{\rm log} \equiv \frac{2 g^2 |I(E_\bos + i \eta_\bos)|}{(2 \pi)^2} \approx 0.057,
\end{equation}
using the numerical value $|I(E_\bos + i \eta_\bos)| = 8.7\times 10^{-4}$ meV$^{-2}$,
validating our weak-coupling assumption.
\SAVE{It will be asked, what is the
relation of this to the standard kinds of dimensionless couplings.
These are $\lambda\equiv 3\int d\omega \alpha^2 F(\omega)/\omega$ in
strong-coupling theory, or $\lambda = {\cal N}(0) |V|$ in BCS theory.
I actually remembered the BCS version as 
$\lambda \propto g^2 {\cal N}(0)/\Omega_0$ but maybe there is
no formal definition of the coefficient there.
Anyhow, it will be noticed that $I(\Ew) \propto {\cal N}(\epsilon)$.
[I am glossing over just what energy is the argument of $\cal N()$, 
it's obviously a range of energies in the integral.]
But if we probe more deeply, using a contour integral over $z$
for energy, and taking the limit of small $\Delta_0$, it seems
we get $I(\Ew) \sim \Delta_0^2 {\cal N}(0) /\Ew^3$.
So long as $\Delta_0$ and $\Omega_0$ are similar in magnitude,
we could write that as $I(\Ew) \sim {\cal N}(0) /E_I$,
where $E_I$ is an energy the order of $\Delta_0$ and $\Omega_0$.}

\SAVE{It is still quite unknown whether the vibration (if it is one)
is of an apical or in-plane oxygen, which determines the symmetry
of the electron-boson coupling.}

\section{Momentum-dependent boson coupling and gap renormalization}

\newcommand{\tg}{\tilde g}
What if the electron-boson coupling $g(\qvec)$ in \eqref{eq:el-boson}
is not constant but 
depends on the electron momentum transfer $\qvec$?
Firstly, it
gives renormalizations of $\Delta(\kvec)$
due to $\Sigma_{12}$ which is no longer zero 
(see Eqs.\eqref{eq:LDOS} and \eqref{eq:freeSCprop}).
To obtain an upper bound for the gap renormalization,
we try the form for $\qvec$ dependence which leads to the maximal renormalization, 
namely $|g(\qvec)|^2 = \tg^2 [\frac{1}{2} (\cos q_x + \cos q_y)]$, where
we set $\tg$ to the fitted $g$ value from Table~\ref{tab:bothfitresults}.
\SAVE{We write $\tg$ to indicate this is not the same as $g$.  
If we use the most naive recipe, namely rms, there still seems to be 
a factor of 2 difference.}
We compute the
gap renormalization using the obvious generalization  
of Eq.~\eqref{eq:selfenergycorr}
to account for $\qvec$ dependence of electron-boson coupling in 
the off-diagonal components of Eq. \ref{eq:selfenergycorr}
(See \ref{subsec:app1} for details).
We find that for all energies, the gap renormalization is
less than 5 meV (See Fig. \ref{fig:appendix_gap_renorm}),  
which is small enough compared to $\Delta_0$
to justify our weak-coupling assumption, but not so small
to categorically rule out some contribution by the boson  
to pairing.

\SAVE{(speculation by SP to be checked):
If the boson propagator has a momentum dependence 
via the mode dispersion $\Omega(\kvec)$, probably the boson feature 
is a superposition of features like in our calculation, convolved with
a distribution of $\Omega$'s...  producing an extended
dip-hump shape?}

For the boson feature, the overall structure of the calculation 
carries through but the self-energy
$\Sigmaa$ becomes momentum dependent.
We find the same sort of \LDOS~ feature, in which 
``$g^2$'' is now interpreted as a certain weighted average of 
$|g(\qvec )|^2$ over the Brillouin zone --  
\SAVE{One certainly cannot discriminate this momentum dependence based
on a single experimental spectrum.}
a lumped parameter in the spirit of the ``$\alpha^2 F(\omega)$'' combination
from the strong-coupling formalism~\cite{McMillan}.
\SAVE{of $s$-wave superconductors}
The singularity in the self energy still come from the
saddle point in the dispersion of the $d$-wave BCS quasiparticles,
leading to the same qualitative shape (smoothed step + logarithm)
for the boson feature . See \ref{subsec:app2} for details.

\SAVE{SP conjectures: If the gap renormalization (say of
energy scale $\delta$) due to the boson at our level of pert theory is much
smaller than an already existing gap (say $\Delta$), then in the absence of 
$\Delta$ that boson can only produce a gap of order $\delta$, though
perhaps with factors of unity (or $4\pi$) that would cloud the
picture. For the present example, it would appear $\delta$ is 
small enough that this won't make a difference.}

\SAVE{Our set-up implicitly assumes this boson
is not the main source of pairing, but we think the calculation is roughly
valid even when it is.}

\SAVE{(From 8/2012)  CLH had an idea for estimating a dimensionless ratio
for the strength of pairing due to $g^2$.  There will be a log
singularity of $\Sigma_d(\Ew)$ at $E_\bos$: at that peak, the
ratio of $\Sigma_d/\Delta_0$ is bigger than it is anywhere else.
Now, from \eqref{eq:selfenergy} we see the only difference in
the $\tauu_1$ part is an extra factor $\Delta(\kvec)/E(\kvec)$,
this becomes practically $\Delta_0/E_\coh$ at the saddle point.
(Well, I need to check if there's an extra factor of $1/2$ from
the $\qvec$ dependence of $|g(\qvec)|2$.)
So the end result looks like \eqref{eq:Sigma_sing} with that
extra factor.  So, $\Sigma_d/\Delta_0 \approx g^2 \meffcoh (\log)/(2\pi)E_\coh$.
The log will be $\ln (E_\bos/\eta)$ roughly, thus something like
$\ln (110/7)\approx 3$.  All in all is (36 meV)$^2$ (94 meV)$^{-1}$ (3)/(6)(40 meV)
= 0.17.  Everywhere else it will be much smaller.
Incidentally $\meffcoh^{-1}$ is roughly proportional to the bare DOS
called ${\cal N}(0)$.}

\section{Conclusion and Discussion}

\SAVE{CLH's outline for the conclusion
(a) summary, including comparison of fitted $\Omega_0$ to other values
(b) competing scheme (Johnson) [CUT except a brief mention intro]
(c) technical point, would boson feature cancel?
(d) big point: estimate d-wave pairing
(e) $\kvec$ dependence as future topic.(TRIMMED TO 1 SENTENCE)}

\SAVE{The data may appear to have multiple humps --
Ref.~\cite{Carbotte2} claimed to see a second boson feature --
so it is pertinent to check how many humps and dips
are engendered by a single boson mode.}

We have shown how a weak-coupling point of view can be used
to analyze the high-energy features in the STM data of BSCCO. 
The ideal analytic shape of the feature is a linear combination of
a (rounded) logarithmic-kink and a (rounded) step edge 
[cf. Eq.~\eqref{eq:deltaN}].
Our proposed fitting scheme allowed us
to extract (1) the boson's frequency $\Omega_0$ (2) an average
electron-boson coupling $g$, and  an estimate
of the damping of the $d$-wave Bogoliubov quasiparticles.
Our estimate $\Omega_0 \approx 56$ meV is in agreement
with previous estimates from STM data, which were not fully 
in agreement with ARPES data \cite{ARPES_1,ARPES_2,ARPES_3,ARPES_4,ARPES_5},
(ARPES results suggest $\Omega_0 \approx 40$ meV.)
\SAVE{see tabulated ARPES refs. in email
thread on 9/14/11, subject  "chapter 4, almost v2 comments.
Cuk et al (Ref.~\onlinecite{ARPES_5})	
refer to an actual ``half-breathing'' O frequency 
of $\sim 70$ meV (but softened by doping) that was measured
by A. Lanzara Ref.~\onlinecite{ARPES_4}.
Make SAVE for the following from SP 7/2013:
APRES papers~\cite{ARPES_1,ARPES_2,ARPES_3,ARPES_4,ARPES_5}
say the energy is 50-80 meV for the observed "kink" in dispersion 
(which is affected
by the real part of some self energy process).
Ref~\onlinecite{ARPES_1}
observed $50 \pm 15$ meV. 
Ref~\onlinecite{ARPES_2}
quoted 70 meV. 
Ref~\onlinecite{ARPES_3}
was not clear.
Ref~\onlinecite{ARPES_4}
quoted 50-80 meV.
SP believes these numbers of $\sim 70$ meV should
have $E_\coh\approx 30$ meV subracted, or maybe
80 meV feature minuse $E_\coh \approx 40$ meV.
Finally, Cuk et al~\cite{ARPES_5}
arguing against an  $E_coh + \Omega$ interpretation since
they see the mode in the normal state too.
They argue that the 40 meV mode is explained by
... the near 40 meV B1g phonon
involving the out-of-plane motion of the in-plane oxygens.}

Our simplified 
simple functional form for the boson feature [Eq. \eqref{eq:bosonfit}]
facilitates the vast number of numerical fits required 
by the extreme spatial inhomogeneity of STM spectra
in BSCCO~\cite{McElroy,inhomogeneity-refs:Pan,inhomogeneity-refs:Cren,inhomogeneity-refs:Howald,inhomogeneity-refs:Lang,inhomogeneity-refs:Fang,Jacob}.
However, our theory did {\it not} address the spatial Fourier
spectrum of the boson feature~\cite{Jinho,Zhu1,Zhu2},
which might distinguish the true functional form of $g(\qvec)$
and thus illuminate the nature of the bosonic mode.

\SAVE{Note $I(\Ew,\eta)$, for a range of $E$ and $\eta$, can be evaluated 
and tabulated once and for all.
A side result of our fits is estimates of the inverse lifetimes $\eta_\coh$ and
$\eta_bos$ of the $d$-wave quasiparticles in both energy windows, 
which gives a consistency check on our using a formalism based on well-defined 
quasiparticles.}


\LATER{1 To be re-examined.  This was one of our
earliest investigations in the boson project.
In the case of an $s$-wave superconductor (i.e. $\Delta(\kvec)$
independent of $\kvec$), and with a dispersionless boson
such as ours, the boson feature would cancel completely
in our weak-coupling treatment,
much as the \LDOS~ signature vanishes for the Kohn 
anomaly of the electron dispersion in normal metals. 
(A glitch is certainly present in the dispersion relation, 
but the increased number of levels in that energy interval is 
exactly canceled by the decreased quasiparticle weight.)
This may have motivated suggestions \cite{Pilgram,Carbotte2} that the observed boson
mode was an artifact localized in a surface barrier,
(rather than in the superconducting CuO$_2$ layer).
We wrote in the 2012 drafts that it followed from
``$\Sigma(\Ew)$ is independent of $\Ew$ for s-wave'' 
but SP is doubtful of that and wanted to check it numerically.
Anyhow, the idea is that the $d$-wave quasiparticle dispersion is
non-uniform enough to make $\Sigmaa(\Ew)$ nonconstant in $\Ew$ 
thus producing an \LDOS~ feature.}


Our approach was agnostic as to the pairing mechanism. 
If the fitted $g$ respects the weak-coupling assumption - as we found
 for a typical spectrum  -
it can be inferred that the boson producing the STM feature is 
{\it not} contributing significantly to the pairing;  if the weak-coupling assumption were to be violated,
we can only conclude that the boson \emph{perhaps} plays a role in the main mechanism.
To resolve that question, one must see if a strong-coupling Eliashberg
calculation predicts a pairing amplitude $\Delta_0$ comparable to the observed  value.


\SAVE{Ref.~\onlinecite{Jinho} performed a Fourier transform
of the spatially-modulated strength of the boson mode
motivated by the prior use of such Fourier transforms on the basic electronic 
\LDOS~ (in the presence of random scatterers) to back out the quasiparticle
dispersion relation around its nodal points~\cite{McElroy}.
Zhu and Balatsky~\cite{Zhu1,Zhu2}, using a model of the same form as our
Eq.~\eqref{eq:HBCS}, numerically evaluated the wavevector dependences
(and identified the best match with experiment 
among their family of models), but gave no analytical justification.
Perhaps an extension of our small-damping theory to the wavevector
dependence will yield a simple picture, in the spirit of the ``octet model''
explaining low-energy modulations in terms of the quasiparticle
dispersion~\cite{McElroy}.  However, in the presence of the well known
disorder of pairing amplitude in BSCCO, it's not clear if this has anything
to do with data.}

\acknowledgments
We thank J. C. Davis, J. W. Alldredge,
D. J. Scalapino, A. Balatsky, P. Hirschfeld, J.-X. Zhu,
and M. Fischer for conversations. We also thank 
J. W. Alldredge for providing us with data sets.
This work was supported by NSF grant DMR-1005466
and CCMR computing facilities.

\appendix
\section{Effects of Momentum Dependent Electron-Phonon Coupling}
\label{sec:appendix}

We quickly recall the basic formula for the weak-coupling self-energy where
we have now an explicitly momentum dependent electron boson coupling $g(\qvec)$
and self-energy $\Sigmaa(\kvec, \Ew)$

\onecolumngrid

\begin{eqnarray}
\label{eq:appendix_selfenergy}
\Sigmaa(\kvec;\Ew) 
& = & \int_\BZ \frac{d^2\qvec}{(2\pi)^2} \frac{g(\qvec)^2}{2} 
\Bigg\{
\frac{(\Ew+\Omega_0) \Id + \epsilon(\kvec - \qvec) \tauu_3 - \Delta(\kvec - \qvec) \tauu_1}{(\Ew + \Omega_0)^2 - E(\kvec - \qvec)^2} 
  + \frac{\Omega_0 ( \Id + \frac{\epsilon(\kvec - \qvec)}{E(\kvec - \qvec)} \tauu_3
- \frac{\Delta(\kvec - \qvec)}{E(\kvec - \qvec)} \tauu_1 )}{[\Ew - E(\kvec - \qvec)]^2-\Omega_0^2} .
\Bigg\}
\end{eqnarray}

\twocolumngrid

As was mentioned in the main text, the two effects of the momentum dependent electron-boson coupling
$g(\mathbf{q})$ are 1) renormalization of the bare d-wave gap $\Delta(\mathbf{k})$, and 2) the self-energy
will no longer momentum independent (as in the main text). 

\subsection{Renormalization of the bare d-wave gap}
\label{subsec:app1}

\begin{figure}
\includegraphics[width=80mm]{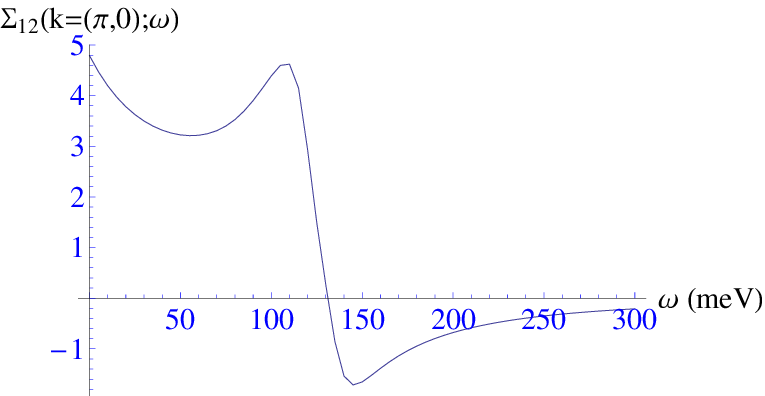}
\caption{In this figure is shown the maximal gap renormalization due to a
momentum dependent electron boson coupling for our chosen band-structure
and extracted boson frequency and electron boson coupling strength. 
We adopt the six-parameter fit from M. R. Norman {\it et al}, Phys. Rev. B 52, 615 (1995):
hopping amplitudes to successive neighbors of 
$t_1=-147.9$ meV, $t_2=40.9$ meV, $t_3=-13.0$ meV, $t_4=- 14.0$ meV
 and $t_5=12.8$ meV, 
plus a chemical potential $\mu = 130.5$ meV as in main text.
\SAVE{To compare with numbers in text, 
$4 c_1=-595.1 meV$, $4 c_2=163.6 meV$, $4 c_3=-51.9 meV$, $8 c_4=- 111.7 meV$
 and $4 c_5=51.0 meV$ (Norman et al used $c$.  The scaled parameters are:
$t_1 = -1$, $t_2 = +0.2749$, $t_3 = -0.0872$, $t_4 = -0.0938$, $t_5 =+0.0857$ and
$\mu = +0.8772$.}
See Table I of the main text for the values of the fitted parameters. We have
plotted for the particular momentum ($\pi$,0). Similar magnitudes are obtained
for other momenta.}
\label{fig:appendix_gap_renorm}
\end{figure}

Consistency of the weak-coupling assumption requires expectedly that the renormalization
of the bare d-wave gap should be at most a non-appreciable fraction of the bare gap.
This is a different consistency check than done in the main text using $\lambda_{log}$. 
$\lambda_{log}$ instead measures the smallness of the \emph{diagonal}
components of the self-energy $\Sigmaa_{11}$ and $\Sigmaa_{22}$ with respect to that
of the diagonal components of the bare (inverse) Green's function.
To estimate an upper bound for the gap renormalization,
we tried the form for $\qvec$ dependence which leads to the maximal renormalization, 
namely $g(\qvec)^2 = \tilde{g}^2 [\frac{1}{2} (\cos q_x + \cos q_y)]$, where
we set $\tilde{g}$ to $50$ meV which is the fitted $g$ plus one error-bar on it.
We computed the
gap renormalization using Eq.~\eqref{eq:selfenergy} which is
\begin{eqnarray}
\label{eq:appendix_selfenergy_12}
\Sigmaa_{12}(\Ew) 
& = & 
 \int_\BZ \frac{d^2\qvec}{(2\pi)^2} \frac{g(\qvec)^2}{2} 
\Bigg\{
\frac{- \Delta(\kvec - \qvec) }{(\Ew + \Omega_0)^2 - E(\kvec-\qvec)^2} \nonumber \\
& & + \frac{\Omega_0 ( 
- \frac{\Delta(\kvec-\qvec)}{E(\kvec-\qvec)}  )}{[\Ew - E(\kvec-\qvec)]^2-\Omega_0^2} .
\Bigg\}
\end{eqnarray}

to account for $\qvec$ dependence of electron-boson coupling in 
(the off-diagonal components of) Eq. \ref{eq:selfenergy}.
We find that for all energies, the gap renormalization is
less than 5 meV as shown in Fig. \ref{fig:appendix_gap_renorm}.
This is small enough compared to $\Delta_0 \approx 40$ meV
to justify our weak-coupling assumption, but not so small
to categorically rule out some contribution by the boson  
to pairing. A small contribution of the observed boson
to the superconductivity that is dominantly established
by the as-yet-unknown mechanism is not unrealistic phenomenologically.

\subsection{Effect on the boson feature}
\label{subsec:app2}

Firstly we recall that for the momentum independent electron boson coupling,
the off-diagonal part of the self-energy is identically zero as mentioned in 
the main text. According to our analysis for the diagonal parts, the singular
contributions are equal for $\Sigma_{11}$ and $\Sigma_{22}$. They are of the form
$(i g^2 a^2 m^*/2\pi) \log [ (\Ew - E_{coh} -
\Omega_0) m^*/K_x K_y]$  + regular terms. 
Here $E(\kvec) \approx E_\coh + (k_x-\ksaddle)^2/2\meffx 
-(k_y-\pi/a)^2/2\meffy $ near its saddle points (here in this
expression assumed to be $\Kvec_* = (\ksaddle,0)$; there are four such saddle points), 
$\meffcoh\equiv \sqrt{\meffx \meffy}$, 
and $K_x$, $K_y$ are cut-offs, 
representing the range of $(k_x,k_y)$ within which this expansion
is valid.

When $g$ has $\kvec$-dependence, the singular
contributions get modified to

\onecolumngrid

\begin{equation}
\Sigma_{11}(\kvec,\Ew) = \Sigma_{22}(\kvec,\Ew) =
\sum_{\Kvec_*}  \frac{i g(\kvec - \Kvec_*)^2 a^2 m^*}{2\pi} \log \left[ \frac{(\Ew - E_{coh} - \Omega_0)m^*}{K_x K_y} \right]
\end{equation}
\begin{equation}
\Sigma_{12}(\kvec,\Ew) = \Sigma_{21}(\kvec,\Ew)^* =
\sum_{\Kvec_*}  \frac{i g(\kvec - \Kvec_*)^2 a^2 m^*}{2\pi} \log \left[ \frac{(\Ew - E_{coh} - \Omega_0)m^*}{K_x K_y} \right]
  \times  \frac{- \Delta(\Kvec_*)}{E(\Kvec_*)}
\end{equation}

\twocolumngrid

where $\Kvec_* = (\pm\ksaddle,0)$ and $(0,\pm,\ksaddle)$
are the saddle-points in $E(\kvec)$ as discussed in the main text.

Now, the off-diagonal part is also non-zero and
$(- \Delta(\Kvec*)/E(\Kvec*))$ is the additional factor for the
off-diagonal terms, and the value of this factor is either +1 or -1
since $\epsilon(\Kvec*)=0$. 
$\Sigma_{12}(\kvec,\Ew)$ has d-wave symmetry as expected.

In the above, we have made the algebraic step that near the saddle point of
the singular denominator of the integrand, rest of the
(regular) terms in the integrand can be replaced by their zero-th
order values. The other terms contribute only to the regular part of the self-energy
(i.e do not appreciably contributed to the qualitative shape of the boson feature).
From the above we see that the off-diagonal terms of self-energy will
have same log singularities as diagonal terms.

Going to shape of boson feature in LDOS,  we get for singular contribution of the electron-boson coupling
to the LDOS :

\onecolumngrid

\begin{align}
\delta n(\Ew \: \text{  near  } \: E_{bos} + i \eta_{bos}) = & \frac{1}{\pi} Im\bigg[ \int_\BZ \frac{d^2\kvec}{(2\pi)^2}
- \Sigma_{22}(\kvec,\Ew) \frac{|\Delta(\kvec)|^2}{ (\Ew^2 - E(\kvec)^2) ^2 }
- \Sigma_{11}(\kvec,\Ew) \frac{ (\Ew + \epsilon(\kvec)) ^2 }{ (\Ew^2 - E(\kvec)^2) ^2 } \nonumber \\
& +
\Sigma_{12}(\kvec,\Ew) \frac{2 \Delta(\kvec) (\Ew + \epsilon(\kvec)) }{ (\Ew^2 - E(\kvec)^2) ^2 } \bigg] \\
= & \frac{1}{\pi} Im\bigg[ \frac{i a^2 m^*}{2\pi} \log \left[ \frac{(\Ew - E_{coh} - \Omega_0)m^*}{K_x K_y} \right]  
( I_1(\Ew)   +  I_2(\Ew)  +  I_3(\Ew) ) \bigg]
\end{align}

\twocolumngrid

where $I_1$, $I_2$ and $I_3 $ come 

from $\Sigma_{11}$ piece :
\begin{equation}
I_1(\Ew) = - \sum_{\Kvec_*} \int_\BZ \frac{d^2\kvec}{(2\pi)^2} g(\kvec - {\Kvec_*})^2  \frac{ (\Ew +
\epsilon(\kvec)) ^2 }{ (\Ew^2 - E(\kvec)^2) ^2 }
\end{equation}

from $\Sigma_{22}$ piece :
\begin{equation}
I_2(\Ew) =  - \sum_{\Kvec_*} \int_\BZ \frac{d^2\kvec}{(2\pi)^2} g(\kvec - {\Kvec_*})^2
\frac{|\Delta(\kvec)|^2}{ (\Ew^2 - E(\kvec)^2)
^2 }
\end{equation}

from $\Sigma_{12}$ piece :

\begin{align}
I_3(\Ew) =  \sum_{\Kvec_*} \int_\BZ \frac{d^2\kvec}{(2\pi)^2} g(\kvec - {\Kvec_*})^2 
\bigg( \frac{-\Delta(\Kvec_*)}{E(\Kvec_*)} \bigg) \times \nonumber \\
\bigg( \frac{2 \Delta(\kvec) (\Ew + \epsilon(\kvec)) }{ (\Ew^2 - E(\kvec)^2) ^2 } \bigg)
\end{align}

As was argued in the main text, the shape of the feature is due to the logarithmic term and thus
is not qualitatively changed due to the momentum dependence of the electron-boson coupling. 
The $I_1$, $I_2$ and $I_3 $
integrals determine the relative contributions of self-energy terms
and they also govern the weighting in the
Brillouin zone averaging of $g(\kvec)^2$. These integrals are reminiscent of the $\alpha^2 F(\Ew)$ term
in Eliashberg theory to represent the electron boson coupling. 
We thus have shown that our assumption of momentum independence is not a bad one 
for extracting a single number $g$ due to the argument elaborated in this
section. This fitted
$g$ is to be interpreted a zone-averaged electron-boson coupling and is a reasonable
estimate of its magnitude.


\begin{thebibliography}{99}
\bibitem{E>0} 
We consider only the $\Ew>0$ part of the spectrum for the fitting scheme;
the same fitting scheme could be applied to the $\Ew<0$ side.
\end{thebibliography}
\end{document}